# AddNet: Deep Neural Networks Using FPGA-Optimized Multipliers

Julian Faraone, Martin Kumm, *Member, IEEE*, Martin Hardieck, Peter Zipf, *Member, IEEE*, Xueyuan Liu, David Boland, and Philip H. W. Leong, *Senior Member, IEEE*

***Abstract*— Low-precision arithmetic operations to accelerate deep-learning applications on field-programmable gate arrays (FPGAs) have been studied extensively, because they offer the potential to save silicon area or increase throughput. However, these benefits come at the cost of a decrease in accuracy. In this article, we demonstrate that reconfigurable constant coefficient multipliers (RCCMs) offer a better alternative for saving the silicon area than utilizing low-precision arithmetic. RCCMs multiply input values by a restricted choice of coefficients using only adders, subtractors, bit shifts, and multiplexers (MUXes), meaning that they can be heavily optimized for FPGAs. We propose a family of RCCMs tailored to FPGA logic elements to ensure their efficient utilization. To minimize information loss from quantization, we then develop novel training techniques that map the possible coefficient representations of the RCCMs to neural network weight parameter distributions. This enables the usage of the RCCMs in hardware, while maintaining high accuracy. We demonstrate the benefits of these techniques using AlexNet, ResNet-18, and ResNet-50 networks. The resulting implementations achieve up to 50% resource savings over traditional 8-bit quantized networks, translating to significant speedups and power savings. Our RCCM with the lowest resource requirements exceeds 6-bit fixed point accuracy, while all other implementations with RCCMs achieve at least similar accuracy to an 8-bit uniformly quantized design, while achieving significant resource savings.**

## I. Introduction

CONVOLUTIONAL neural networks (CNNs) have been widely adopted in recent computer vision applications due to their superior prediction capabilities, with researchers gravitating toward larger networks with higher computational complexity and memory requirements [1]. Field-programmable gate array (FPGA) implementations have demonstrated improved latency and power efficiency compared with central processing unit (CPU) and graphics processing unit (GPU) technologies (see [2] and [3]). In contrast to CPU/GPU technologies, they allow customized data paths, enabling improved parallelism and less data movement. This design flexibility poses an opportunity to optimize system performance through custom hardware tailored to the application.

Optimizations via compression, quantization, and neural network layer explorations have been utilized to reduce complexity and boost performance (see [4] and [5]). In particular, quantizing inference networks to very low precision, such as constraining weight representations to binary or ternary values, both reduces memory requirements and enables multiplications to be replaced with the exclusive NOR operation [3], [6]. However, the disadvantage of extreme quantization is that the networks typically incur significant accuracy degradation for very low precisions, especially for complex problems.

One limitation with traditional fixed-point quantization is that it has a uniform distribution. However, it has been demonstrated that a nonuniform distribution with the same number of potential weights can result in better accuracy, provided the distribution appropriately matches the desired full-precision neural network weight distribution [7], [8]. It follows that reducing precision may not be the best method to save silicon area. Reconfigurable constant coefficient multipliers (RCCMs) are an alternative method to reduce FPGA resources through time multiplexing and resource sharing [9]. They are usually realized using additions, subtractions, bit shifts, and multiplexers (MUXes), meaning that multiplies are implemented without requiring digital signal processing (DSP) blocks on an FPGA. However, RCCMs are restricted to a given number of target coefficients; this has restricted their use to DSP application domains including digital filtering and linear transformations (see [10]). We propose a method, AddNet, to design RCCMs with coefficient sets that approximate the desired distribution of neural network weights. Furthermore, we develop a method to train neural networks to take advantage of RCCMs. In doing so, we demonstrate that using AddNet to optimize neural networks outperforms low-precision arithmetic in terms of accuracy for a given silicon area budget.

AddNet consists of the following stages. First, we design a family of RCCMs which are customized to the underlying

Manuscript received February 12, 2019; revised June 5, 2019; accepted August 11, 2019. This work was supported in part by the Australia–Germany Joint Research Co-operation Scheme and in part by the German Academic Exchange Service (DAAD) under Grant 57388068. *(Corresponding author: Julian Faraone.)*

J. Faraone, X. Liu, D. Boland, and P. H. W. Leong are with the School of Electrical and Information Engineering, The University of Sydney, Sydney, NSW 2006, Australia (e-mail: julian.faraone@sydney.edu.au; david.boland@sydney.edu.au; philip.leong@sydney.edu.au).

M. Kumm is with the Applied Computer Science Faculty, Fulda University of Applied Sciences, 36037 Fulda, Germany (e-mail: martin.kumm@cs.hs-fulda.de).

M. Hardieck and P. Zipf are with the Digital Technology Group, University of Kassel, 34121 Kassel, Germany (e-mail: hardieck@uni-kassel.de; zipf@uni-kassel.de).

logic elements on the FPGA. These exhibit very low resource usage and have varying coefficient sets. The RCCM coefficient set whose distribution best replicates the weight distribution of a pretrained network is chosen and the network is retrained with weights restricted to these coefficients. This allows the optimizer to update network weight parameters during training while incorporating information about the underlying hardware. This study does not consider the embedded multipliers present in all modern FPGAs; in practical implementations, we envisage different CNN layers using embedded multipliers or our RCCM, depending on resource and throughput requirements.

The trained network is able to learn a representation compatible with the underlying optimized RCCM, achieving both high performance and accuracy. This allows a significant reduction in resource usage for a given throughput, making our designs suitable for resource-constrained implementations. Additionally, we can scale the parallelism of the design to achieve much higher frame rates for similar resource usages. Specifically, this article makes the following contributions.

1) A novel family of arithmetic RCCM circuits tailored to the FPGA fabric for neural network applications which significantly reduces resource requirements.
2) A distribution matching technique which allows a specific RCCM to be selected based on the required distribution of weights in a CNN and a training algorithm which finds solutions compatible with the selected RCCM.
3) We demonstrate that our method achieves significant improvement in accuracy over low-precision (1–6 bit) implementations and significant reductions in lookup table (LUT) usage over 8-bit fixed-point precision with no loss in accuracy for state-of-the-art networks such as ResNet [11] implemented in fixed point. Moreover, weight storage requirements are reduced through implicit weight sharing.

The remainder of this article is structured as follows. Section II provides a background to training CNNs and constant coefficient multipliers. In Section III, recent state-of-the-art research on quantization training and hardware architectures for the implementation of CNNs is reviewed. Our methodology for designing our RCCMs is described in Section IV. Our training techniques and selection of RCCM are presented in Section V. The hardware architecture used for evaluating the effects of our methods is described in Section VI, followed by training and resource usage results in Section VII. Finally, we conclude this article in Section VIII.

## II. Background

In this section, we discuss the basic computation of CNNs and introduce fixed-point training and softcore multiplier optimizations.

### A. Convolutional Neural Networks

CNNs are biologically inspired networks that process input tensors (multidimensional arrays) of $(w, h, d)$ dimensions in a translationally invariant manner [1]. Typically, $w$ and $h$ are spatial dimensions and $d$ is the number of channels. Processing operations, such as convolution, pooling, and activation functions, are applied in a series of layers, each of which transforms the input tensors from dimensions $(w, h, d) \rightarrow (w', h', d')$. A convolutional layer produces $d'$ output channels, each formed through a convolution of the input tensor with a $J_l \times K_l$ convolutional kernel window, where typically $J_l, K_l \ll w, h$, so it operates on local input regions. The output of convolutional layer $l$ takes as input $S_l$ images of spatial dimensions $w_l$ and $h_l$ and $d$ is the number of neurons. The pixel $y_{l,d,w,h}$ at location $(w, h)$ for the $d$th neuron is calculated as

$$y_{l,w',h',d'} = g\left(\sum_{s=0}^{S_j}\sum_{j=0}^{J_l}\sum_{k=0}^{K_l} w_{l,d',s,j,k} \cdot y_{l-1,d,w+j,h+k}\right) \quad (1)$$

where $g$ is the elementwise activation function [such as the rectified linear unit (ReLU) function $g(x) = \max(0, x)$]. The outputs of a layer are used as inputs to the next layer. A 2-D convolutional layer can be described as matrix multiplication, followed by the elementwise activation function. Similarly, the output of a fully connected layer can be described as

$$\mathbf{y} = g(\boldsymbol{W}^T \boldsymbol{x}) \quad (2)$$

where $\boldsymbol{x} \in \mathbb{R}^{d\times w\times h}$ is the input, $\boldsymbol{y} \in \mathbb{R}^{d'\times w'\times h'}$ is the output, and $\boldsymbol{W} \in \mathbb{R}^{d'\times d\times w\times h}$ are the weights of a linear transformation. Convolutional layers form the bottleneck for CNN implementations, and this tensor form allows efficient matrix-multiplication libraries to be applied.

Pooling layers are downsamplers of 2-D images. Max pooling layers provide a spatial maximum function, which divides an input image into small subtiles of a given window size and then replaces these with the maximum value in the subtile. An average pooling layer is similar; however, it finds the average in the subtile rather than the maximum.

### B. Fixed-Point Training

Deep neural network (DNN) training is an iterative process which has a feedforward path to compute the output and a backpropagation path for learning, which involves calculating gradients and update the network weights. Training of low-precision networks typically involves maintaining a set of single- or double-precision floating point weights $W$ which are quantized to a representation $q$ prior to inference (see [6]). As the quantization functions employed are piecewise and constant, the gradients of quantized weights can be calculated and applied to update their corresponding full-precision weights [12]. A quantization function which reduces mismatch in forward and backward paths is crucial for high accuracy.

To further improve accuracy, an alternative to low-precision networks is to use a weight-sharing approach [13], [14]. Weight sharing involves choosing a finite set of full-precision weights indexed by a codebook. Typically, these weights are chosen to match the desired distribution to reduce information loss, unlike traditional fixed-point quantization where weights are uniformly distributed. Keeping the number of different weights in the codebook small reduces the word size of the indices leading to a small memory footprint. However, weight sharing is normally not applied in FPGA implementations as the weight mapping process introduces additional delays in the critical path of the circuit and requires extra hardware. Furthermore, higher precision arithmetic units also consume more area. For the proposed RCCM, an implicit weight sharing is utilized, reducing coefficient memory without requiring any mapping hardware. Meanwhile, our RCCMs are optimized for FPGA hardware, meaning that they consume less area than fixed-point equivalents.

### C. Small Softcore Multipliers

Due to the low-precision requirements of neural networks, efficient implementations of small multipliers recently have gained growing interest [15], [16]. As FPGAs provide embedded multipliers, it seems natural to use them. For small multiplications, there is a way to perform two multiplications up to $8 \times 8$ bit in a single DSP of typically 18 bits [16]. In case the embedded multipliers are not sufficient, efficient logic-based (i.e., softcore) multiplier implementations are necessary. The use of radix-4 Booth encoding together with an FPGA mapping that maps both Booth encoder and decoders in the same LUT showed to be the most efficient way to implement softcore multipliers leading to up to 50% resource reductions [17], [18] on Xilinx FPGAs. Unfortunately, they are only this efficient for large word sizes of 16 bits and above. For lower word sizes, Xilinx Coregen showed the best results [18]. An optimization which is particularly suited for small multipliers by restructuring common multiplier algorithms was recently proposed in [15]. They were optimized for Intel Stratix 10 showing the smallest resources and latency. The optimizations described in our work add further constraints designing multipliers which do not allow arbitrary fixed-point number support. This is achieved by applying concepts from reconfigurable multipliers.

## III. Related Work

Many quantization methods for neural networks have been explored with the aim of achieving efficient inference in hardware. An efficient way of training networks with different forward and backward functions was introduced in [12]. This led to new derivations of uniform quantization functions for low-precision neural networks in [19] and [20]. Learning symmetric quantization (SYQ) [21] further explored the importance of initializations and designing a quantization function which reduces the forward and backward mismatches. They achieved state-of-the-art accuracies under low-precision weights and activations. This inspired the derivation of the distribution matching initialization method for efficient quantization. Effective nonuniform quantization forms were also explored in the form of log representations [7]. This form can also compute multiplier-less multiply and accumulate (MACs) operations; however, the distribution of the representations is restricted to the log domain.

There have been several accelerator architecture designs for low-precision CNNs with uniform quantization arithmetic. Recent literature includes commercial architectures [22], [23] and also academic approaches [24]–[28]. The benefits, in terms of power and throughput, of fitting a design on-chip was described in [3]. Other FPGA architectures have been implemented to utilize the highly amenable nature of CNNs which constrain weight parameters to be only binary or ternary representations [29], [30]. With restrictions in the efficiency of both software and hardware implementations of neural networks, software-hardware codesign is considered an effective approach to achieve optimal performance [31], [32]. A method for designing a quantization function for both increasing accuracy of binarized CNNs while maintaining efficient multiplier-less hardware was proposed in [33]. In addition, an efficient long short-term memory (LSTM) implementation in [34] utilized load-balance-aware pruning to achieve both network compression and high hardware utilization. Similarly, training highly sparse ternary networks and designing efficient CNN hardware for exploitation was described in [30]. To the best of our knowledge, AddNet is the first quantization scheme which embeds reconfigurability directly into its representations.

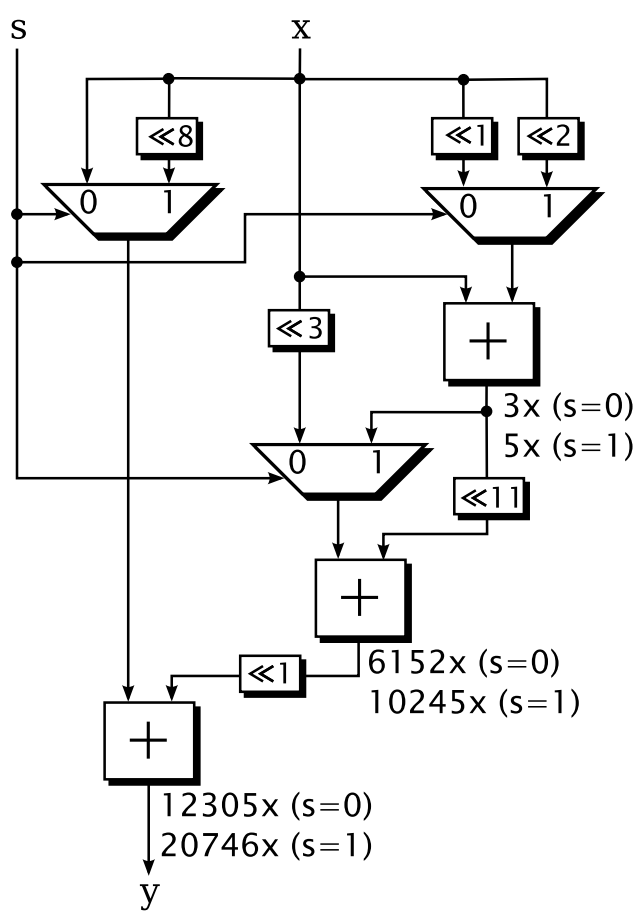

Fig. 1. Example of a reconfigurable multiplier with the coefficient set {12305, 20746}.

## IV. AddNet Reconfigurable Multipliers

In this section, we introduce reconfigurable multipliers and describe their design in AddNet.

### A. Reconfigurable Multipliers

A constant coefficient multiplier is a circuit which computes $y = cx$, using only additions, subtractions, and bit shifts, where $c$ is some predefined number. For example, to compute $y = 6x$ in terms of additions and shifts, we can use

$$(x << 2) + (x << 1) = 6x. \tag{3}$$

The "$<<$" operator represents an arithmetic left shift.

An RCCM is a circuit which computes $y = c_s x$, where $c_s$ is an element from a discrete coefficient set $C = \{c_0, c_1, \ldots, c_{N-1}\}$, chosen from a $\lceil \log_2(N) \rceil$ bit select signal $s$ [35]. RCCMs are usually realized using additions, subtractions, bit shifts, and MUXes. Previous work has shown potential for reducing resource usage compared with a generic multiplier, especially for small values of $N$ [9], [35]–[37].

Fig. 1 shows an example of an RCCM with coefficient set $C = \{12305, 20746\}$. In this example, there is one

2:1 MUX for each adder, each having $s$ as the select line input. The three adders sum various shifted versions of $x$. Each adder is assigned a coefficient set where the value of each row corresponds to the multiple for each configuration. For instance, the top-most adder computes

$$\begin{cases} x + (x << 1) = 3x, & \text{if } s = 0 \\ x + (x << 2) = 5x, & \text{if } s = 1. \end{cases}$$

The bottom-most adder outputs the final output $y$ with coefficient set $C = \{c_0, c_1\} = \{12305, 20746\}$ multiplier-less by

$$y = \begin{cases} x + ((x << 3) + ((x + x << 1), & \text{if } s = 0 \\ \quad << 11) << 1) = 12305x & \\ (x << 8) + ((x + x << 2) + ((x + x << 2), & \text{if } s = 1 \\ \quad << 11) << 1) = 20746x. & \end{cases}$$

By utilizing MUXes in this way, the computation of $c_1 = 12305$ is able to reuse the adders from computing $c_2 = 20746$ and vice versa.

To date, prior research with RCCMs has focused on the design of an RCCM for a predefined set of target constants (e.g., obtained from a digital filter design). This design using minimal resources is an NP-complete optimization problem [36]. However, we want to use RCCMs in neural networks, where the coefficients (weights) are not known in advance. As a result, we invert the RCCM design, and instead of searching for an RCCM circuit for a given coefficient set, this article aims to find one with very low resource usage and a maximum of "useful" coefficients. This low-cost RCCM then replaces multipliers in a conventional CNN implementation. Instead of storing the coefficients, the corresponding select values are stored, which also has the side effect that it requires fewer bits of storage than the direct coefficient value.

### B. FPGA Multiplier Mapping

We searched for building blocks that efficiently map to the logic fabric of an FPGA. Our designs are optimized for the latest Xilinx FPGAs (Virtex 5+6, the seventh generation FPGAs and UltraScale/UltraScale+ FPGAs), but similar circuits can be found for other FPGAs. For these devices, a slice provides either six-input LUTs with a single output (used in Topology A) or two five-input LUTs with shared inputs (used in Topology B). As such, we designed our base topologies to ensure the MUXes fit into the same LUTs that are required for the adders.

Fig. 2 shows the two base topologies used to build the RCCM units in this article. Each of these consists of an adder with at least one input being the output of a MUX. These topologies allow operations of the form $\pm A_p \pm B_q$. For Topology A, $A_p$ can consist of up to four different input values ($p \in 1, .., 4$) with $A_4 = 0$ and $B_q$ can only take one value, that is, ($q = 1$). For Topology B, $p \in 1, .., 3$ with $A_3 = 0$ and $q \in 1, 2$, with $B_2 = 0$. The sign and source signals are selected using a 2-bit input signal $s$. Since there are more possibilities than MUX inputs, a function $\sigma(s)$ is used to choose the actual operation, where $\sigma(s)$ is determined at the design time but may be different for each individual RCCM. Note that there is another possibility to map more input sources to the adder as described in [38]; however, to ensure the topology fits into a single LUT, this comes at a cost of less select inputs. Through our experimentation, we found that the chosen topologies were sufficient for creating RCCMs with a desired coefficient set to simplify the training process. This is further described in Section V.

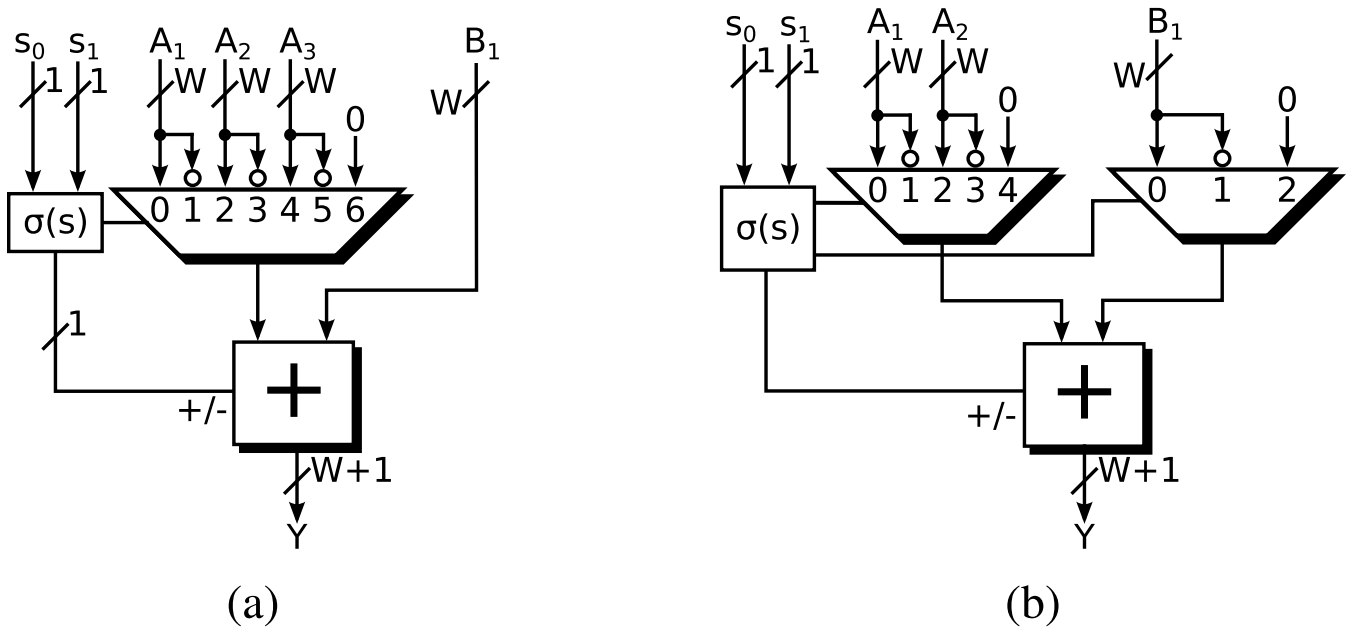

Fig. 2. Base topologies used to build reconfigurable multipliers. (a) Topology A. (b) Topology B.

All contemporary FPGA devices are similar in that their logic blocks consist of LUTs followed by a fast carry chain. Hence, a simple adder can be extended by MUXes with no additional cost for certain MUX sizes when carefully selected for the target device. The detailed slice mappings of our base topologies are shown in Fig. 3, highlighting how our design consumes exactly the same silicon area as a traditional ripple-carry adder with the same word size on that FPGA (which would only implement the XOR gate to complete the carry logic to a full adder).

### C. Architectures Considered

The base topologies described previously can be combined in many ways to design RCCM units. Topology A has the advantage of a potentially larger coefficient set as it allows three different sources at input $A_p$. On the other hand, Topology B has the property that input $B_q$ can be negated or zeroed, which provides symmetric coefficients around zero (as $A_p - B_1 = -(-A_p + B_1)$). We designed three different RCCM architectures from these topologies shown in Fig. 4. These consist of one to three elements of Topology A in the early stages and Topology B at the output stage to ensure symmetric coefficients. Symmetric coefficients improve the ability to match the distribution of the coefficient sets to the pretrained neural network weights which are typically and also approximately symmetric around zero. The benefits of this are further discussed in Section V. Note also that these designs can be trivially pipelined.

As shown in Fig. 4, the $A_p$ inputs to the topologies are all connected to left shift operations $\varphi_{ij}$, which are all hard-wired, since these do not require any LUT resources. It follows that the supported coefficient set depends on the operation mapping function $\sigma(s)$ and the fixed bit shifts $\varphi_{ij}$. As mentioned in Section IV-B, each instance of base Topology A or B consumes the same area as a traditional ripple-carry adder. Hence, as the RCCMs of Fig. 4 consist of two, three, and four base topologies, they are, respectively, called 2-Add, 3-Add, and 4-Add RCCMs in the following.

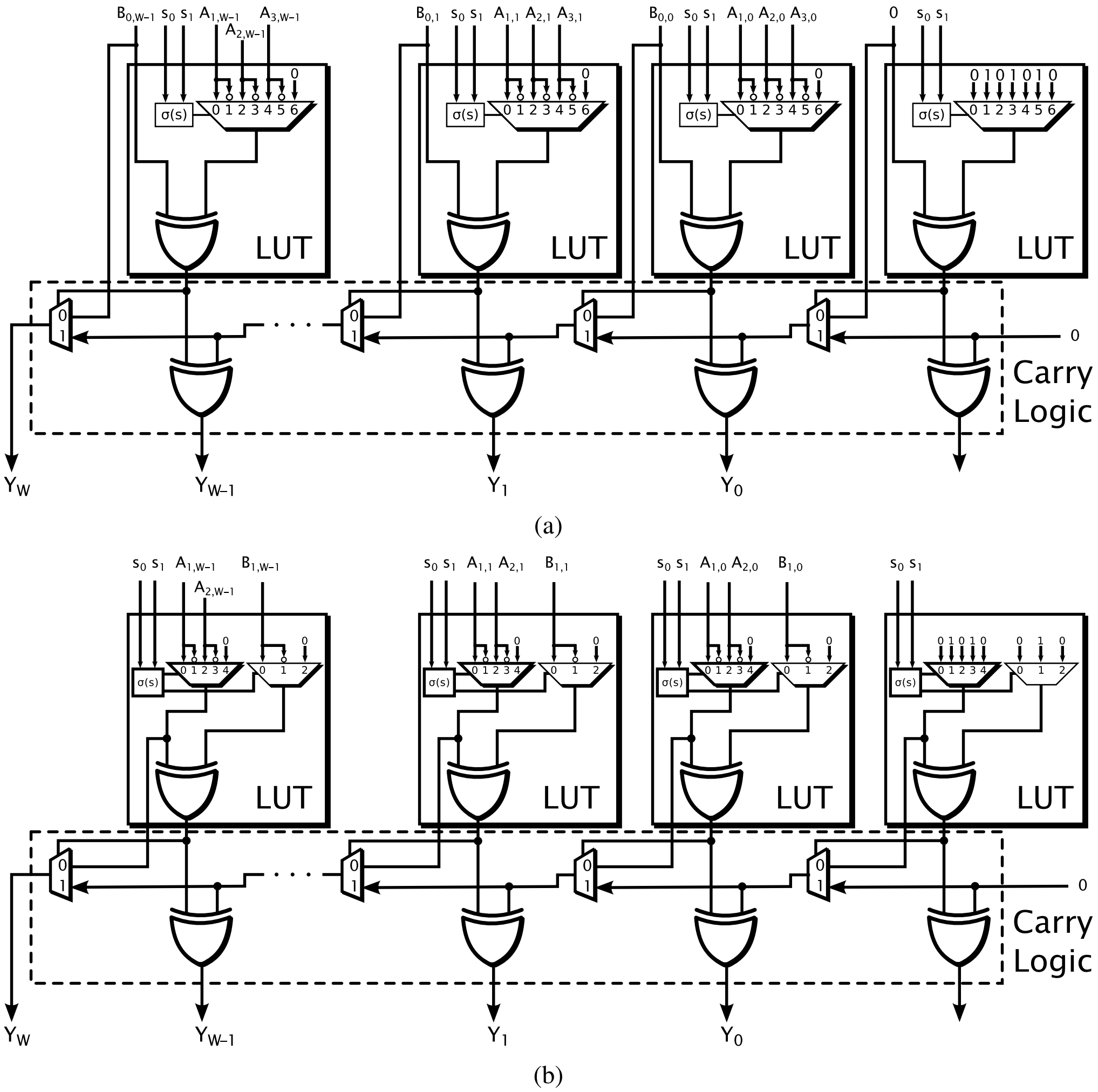

Fig. 3. Bit-level FPGA slice mapping of base topologies of Fig. 2. This is applicable to any FPGA using six-input LUTs, including Xilinx Ultrascale and Intel Stratix X devices. (a) Topology A. (b) Topology B.

The obtained RCCM architectures can multiply with up to $2^{w_s}$ different coefficients, where $w_s$ denotes the total number of bits used for the select signal. For the 2-Add, 3-Add, and 4-Add RCCMs, this translates to $w_s = 4, 6$, and 8, respectively, as shown in Fig. 4. We chose to evaluate coefficient sets where Topology A had four different mapping functions and Topology B had a single one. In addition, all maximum bit shifts were set to $\varphi_{\max} = 3$. This limits the total number of unique combinations, as shown in Table I. With these coefficient sets, an exhaustive enumeration of possible coefficient combinations is feasible with a few minutes of computation time. This allows us to then find the desired coefficient set based on its similarity to the pretrained neural network weight distribution. We note that it may be possible to improve on our results by exploring more mapping functions, which would generate a larger number of unique coefficient sets but at the cost of longer execution time.

## V. AddNet Training

Section IV described a family of optimized multipliers. In this section, we now address the issue of finding the best coefficient set for a given neural network. Neural networks can typically tolerate a certain amount of regularization for their weight representations before the accuracy is impinged upon. Thus, our strategy is to utilize this knowledge and select an RCCM coefficient set which exhibits a distribution similar to the distributions of the neural network weights and retrain the network to learn the representation of the coefficient set.

TABLE I

Properties of the Evaluation of the Proposed RCCM Units With Maximum Possible Set Size $S = 2^{w_s}$

| RCCM | $w_s$ | #unique coefficient sets |
|---|---|---|
| 2-Add | 4 | 1145 |
| 3-Add | 6 | 44198 |
| 4-Add | 8 | 4040952 |

### A. Distribution Matching

To achieve high accuracy in quantized neural network training, it is important to reduce quantization error by using a function which can efficiently map its representations to the full-precision values. This is important to minimize information loss and to achieve a good initialization for training [21]. Fixed-point representations using quantization typically uniformly partition the weight parameter space. However, representations using RCCM coefficient sets discussed in Section IV are nonuniformly partitioned and can vary in size, range, and the nature of the distribution. Thus, for efficient training, we choose an RCCM with a coefficient set to match the distribution of a pretrained model. We use the Kullback–Leibler divergence [39] as a measure of the similarity of two distributions. Let $R$ denote the distribution of the coefficient set of the RCCM, $P$ is the reference distribution of the pretrained model weights, and $N$ is the total number of weights. The Kullback–Leibler divergence $D_{\text{KL}}$ is defined as

$$D_{\text{KL}}(P\|R) = \sum_{i=0}^{N-1} P(i)\, \log \frac{P(i)}{R(i)}. \tag{4}$$

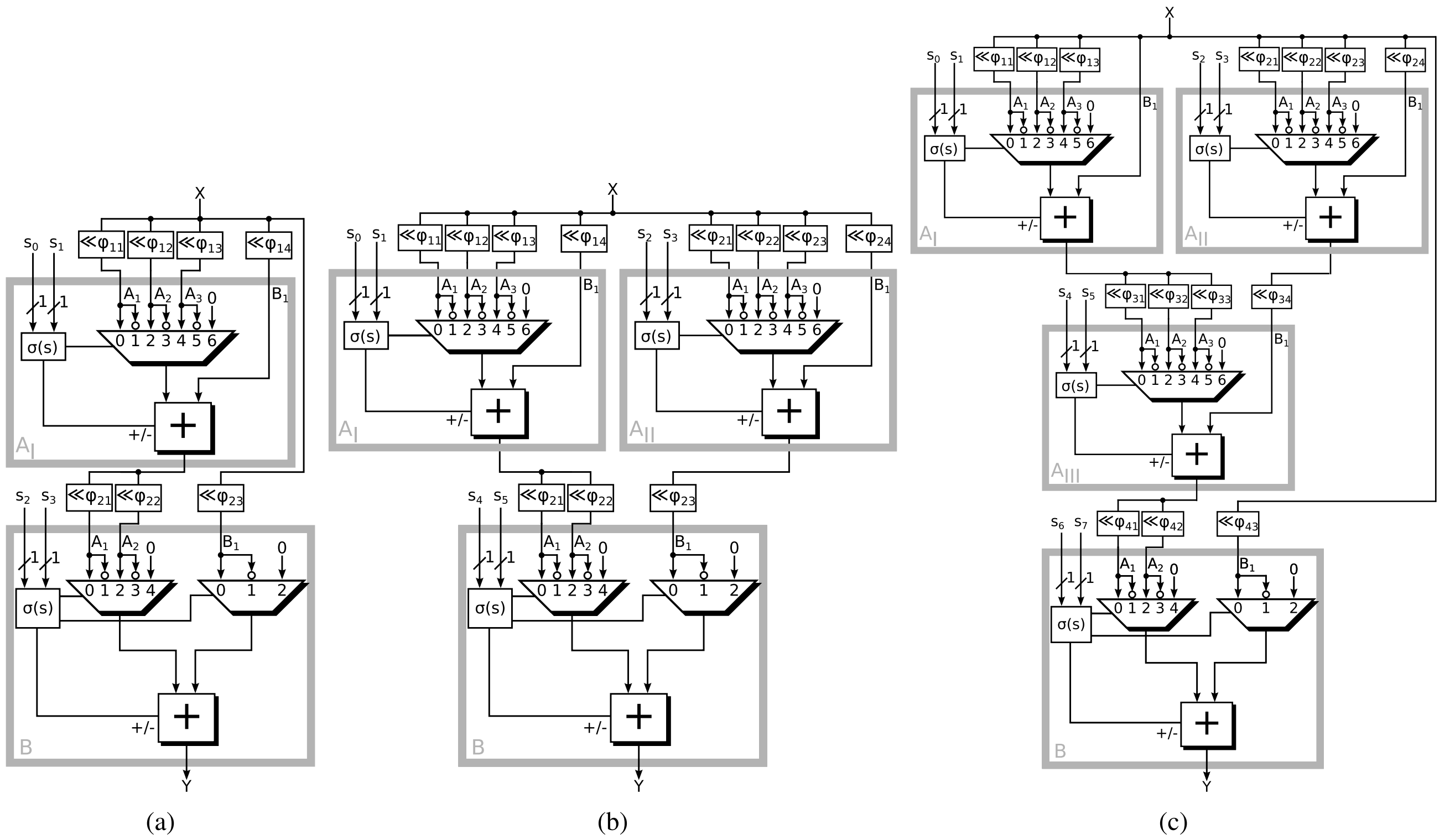

Fig. 4. Selected RCCM circuits. (a) 2-Add RCCM. (b) 3-Add RCCM. (c) 4-Add RCCM.

Thus, for each enumeration of the coefficient sets, we measured the divergence $D_{\text{KL}}$ to the pretrained network weights and selected the top-5 sets with the smallest divergence. We call this technique distribution matching. From the top-5 sets, we selected the set with the largest number of coefficients, to maximize the number of representable states for the weights during retraining. As a secondary criterion, we only selected coefficient sets that include zero. Note that a zero weight could also be alternatively realized by resetting the output flip-flop in a pipelined implementation. Since this leads to an additional select bit that has to be stored in the coefficient memory or a separate decoder this was not further investigated.

To give an example, the weight distribution (using 31 bins) from AlexNet on ImageNet is given in Fig. 5(a). As shown, the weight parameters in this example follow a distribution similar to a Gaussian distribution, meaning that small weight values near zero occur much more often than large values. The coefficient sets of the RCCM circuits of Fig. 4 with the best distribution matching are given in Table II, with their corresponding configuration parameters in Table III. Their distributions are shown in Fig. 5(b)–(d) which are similar to the pretrained model. We call these optimized 2-Add, 3-Add, and 4-Add RCCM circuits.

The exhaustive search for coefficient combinations yields the distributions of different natures, meaning that this method would most likely be able to efficiently map to other potential network weight distributions. To further justify our approach of distribution matching, we also study an RCCM with an unoptimized choice of coefficient set with differing distribution nature. Fig. 5(e) shows the distribution with the worst (i.e., largest $D_{\text{KL}}$) divergence score for 63 coefficients. It is not obvious that the corresponding coefficient set, $C = \{0$ 8 12 14 16 18 20 21 23 24 36 38 40 42 44 45 47 49 51 52 54 56 58 60 68 70 72 74 76 77 79 88$\}$, would lead to poor CNN inference accuracy. However, as shown in Table IV, when used with AlexNet [40] in the 2-Add case, Top-1/Top-5 accuracy is 53.8%/76.9% (results are presented as a Top-$k$ percentage, where a classification is considered correct if the actual class is among the highest $k$ probabilities). With distribution matching, the accuracy is 55.8%/79.8%, which is significantly better than the unoptimized set and equivalent to full-precision floating point accuracy.

### *B. Weight Quantization*

To both exploit our RCCM and achieve high accuracy, our network should be trained to match the underlying inference

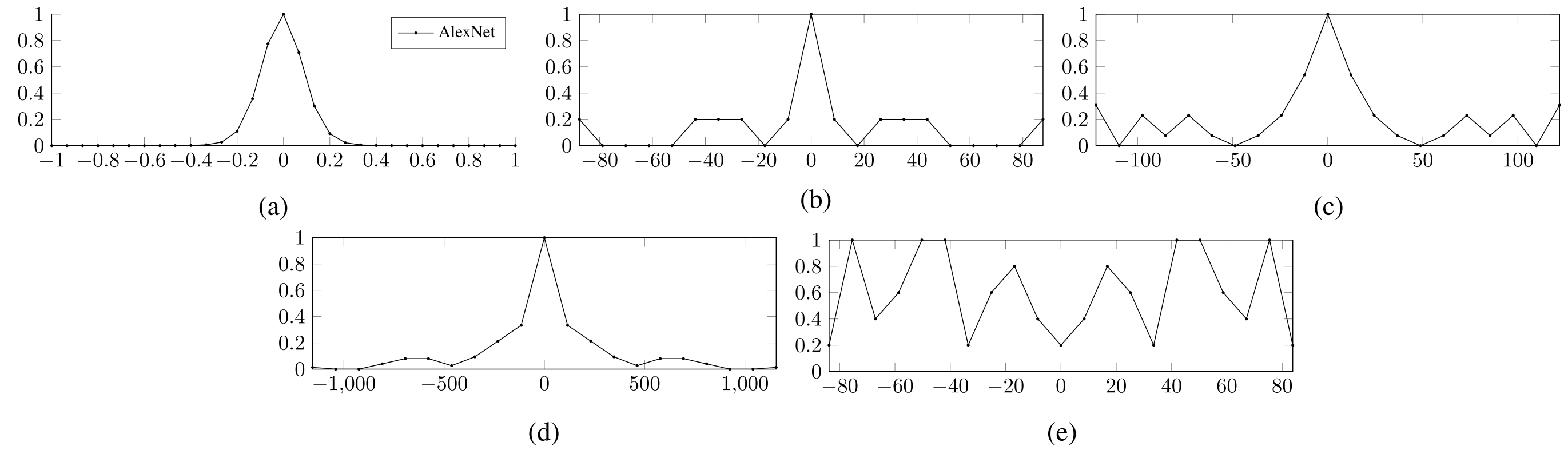

Fig. 5. Distribution for CNN weights and constant multiplier coefficients. (a) Weights of AlexNet. (b) 2-Add—15 coefficients and optimized. (c) 3-Add—59 coefficients and optimized. (d) 4-Add—207 coefficients and optimized. (e) 3-Add—63 coefficients and unoptimized.

TABLE II
OPTIMIZED RCCM COEFFICIENTS

| arch. | #coeff | Coefficient set (±) |
|---|---|---|
| 2-Add | 15 | 0 1 2 8 28 36 44 92 |
| 3-Add | 59 | 0 1 2 3 4 5 6 7 9 10 12 13 14 16 23 29 30 32 63 69 70 72 87 93 94 96 119 125 126 128 |
| 4-Add | 207 | 0 1 2 4 5 7 8 9 11 13 14 15 16 18 19 20 21 22 23 24 25 26 27 28 29 30 31 32 33 34 36 37 38 39 40 46 48 54 58 64 69 70 71 74 75 76 78 80 81 82 84 85 87 94 96 102 114 118 126 134 142 150 166 174 182 190 194 198 206 214 222 230 238 246 258 262 270 278 286 302 310 318 326 334 382 398 446 450 526 566 574 582 614 622 654 662 670 686 694 710 766 782 830 1214 |

TABLE III
CONFIGURATION PARAMETERS OF THE RCCM UNITS

| RCCM | type | $s_1\ s_0$ | | | | shifts | | | |
|---|---|---|---|---|---|---|---|---|---|
| | | 00 | 01 | 10 | 11 | $\varphi_{i1}$ | $\varphi_{i2}$ | $\varphi_{i3}$ | $\varphi_{i4}$ |
| 2-Add | $A_I$ | $A1$+$B1$ | $A2$+$B1$ | $A3$+$B1$ | $B1$ | 0 | 1 | 3 | 2 |
| | B | -$A1$+$B1$ | -$A2$+$B1$ | $A1$-$B1$ | $A2$-$B1$ | 0 | 3 | 2 | – |
| 3-Add | $A_I$ | $A1$+$B1$ | $A2$+$B1$ | $A3$+$B1$ | -$A2$+$B1$ | 0 | 2 | 3 | 3 |
| | $A_{II}$ | $A1$+$B1$ | $A2$+$B1$ | $A3$+$B1$ | -$A1$+$B1$ | 0 | 1 | 3 | 0 |
| | B | -$A1$+$B1$ | -$A2$+$B1$ | $A1$-$B1$ | $A2$-$B1$ | 0 | 3 | 0 | – |
| 4-Add | $A_I$ | $A1$+$B1$ | $A2$+$B1$ | $A3$+$B1$ | -$A3$+$B1$ | 0 | 1 | 3 | 0 |
| | $A_{II}$ | $A1$+$B1$ | $A2$+$B1$ | $A3$+$B1$ | -$A2$+$B1$ | 0 | 1 | 3 | 1 |
| | $A_{III}$ | $A1$+$B1$ | $A2$+$B1$ | $A3$+$B1$ | -$A1$+$B1$ | 0 | 1 | 3 | 3 |
| | B | -$A1$+$B1$ | -$A2$+$B1$ | $A1$-$B1$ | $A2$-$B1$ | 0 | 3 | 1 | – |

hardware. During our fixed-point training, for each layer $l$, we first clip the weights so that $w_l \in (-M, M)$, where $M$ is a range hyperparameter at each inference step and then quantize them to fixed-point representations. As discussed in Section IV, our multiplier consists of a fixed-point input and a value from $C$. During AddNet training, we introduce

TABLE IV
ACCURACY CHANGE FROM OPTIMIZED DISTRIBUTION MATCHING ON ALEXNET FOR THE 2-ADD CASE

| | Unoptimized | Distribution Matching | 32bit Float. |
|---|---|---|---|
| Top-1 | 53.8% | **55.8%** | 55.1% |
| Top-5 | 76.9% | **79.8%** | 79.2% |

a function whereby every floating point weight is quantized according to

$$q(w_l) = \arg\min_{c_i \in C'} |c_i - |w_l|| \tag{5}$$

where $c_i \in C'$ represents the possible positive coefficients of $C$ scaled by $\lambda_l$. Here, (5) aims to minimize the quantization error between the quantized weight values and the representations in our coefficient set. The scaling with $\lambda_l$ is done so that $q_i \in (-M, M)$ where $M$ is initially the range of the pretrained model. By using distribution matching, we minimize this quantization error to achieve an efficient initialization. We then retrain the network using the straight through estimator (STE) approach as described in [12]. This approach allows a nondifferentiable function defined in the forward path to use a nonzero surrogate derivative function in the backward-path gradient calculations. Thus, in our case, we allow

$$\frac{\partial L}{\partial q} = \frac{\partial L}{\partial w} \tag{6}$$

where $L$ is the loss function. The quantized weights $q(w_l)$ are used for inference in the forward path, and the floating point weights $w_l$ are updated in the backward path. During training, $\lambda_l$ becomes a parameter which is also updated during backpropagation. By using a representation compatible with the multiplier in the forward path, the network learns a representation both high in accuracy and hardware efficiency. After training, the floating point weights are discarded and $q(w_l)$ is used for hardware deployment.

### *C. Activation Quantization*

As initial training results did not show accuracy degradations compared with activations larger than 8-bit 2's complement, we first uniformly quantize the activations to 8 bits.

**Algorithm 1** Training a CNN Using AddNet Representations

**Initialize:** Pre-train model
Set adder size
c = DistributionMatching($\sigma(s)$) using (4)
**Inputs:** Minibatch of inputs & targets $(I, Y)$, Loss function $L(Y, \hat{Y})$, current weights $\boldsymbol{W_t}$ and learning rate, $\gamma_t$
**Outputs:** Updated $\boldsymbol{W_{t+1}}$, $\boldsymbol{\lambda_{t+1}}$ and $\gamma_{t+1}$

***Forward propagation:***
**for** l=1 to L **do**
  $\boldsymbol{Q}_l$ = Quantize($\boldsymbol{W}_l$) using (5) and (7)
**end for**
$\hat{Y}$ = **ForwardPropagation** $(I, Y, \boldsymbol{Q}_l)$ using (7)

***Backward Propagation:***
$\frac{\partial \hat{L}}{\partial \boldsymbol{Q}_l}$ = **WeightBackward**($\boldsymbol{Q}_l$, $\frac{\partial \hat{L}}{\partial \hat{Y}}$)
$\frac{\partial \hat{L}}{\partial \boldsymbol{\lambda}_l}$ = **ScalarBackward**($\frac{\partial \hat{L}}{\partial \boldsymbol{Q}_l}, \boldsymbol{\lambda}_l, \frac{\partial \hat{L}}{\partial \hat{Y}}$)
$\boldsymbol{W_{t+1}}$ = **UpdateWeights**($\boldsymbol{W_t}$, $\frac{\partial \hat{L}}{\partial \boldsymbol{Q}_l}, \gamma$)
$\boldsymbol{\lambda_{t+1}}$ = **UpdateScalars**($\boldsymbol{\lambda_t}$, $\frac{\partial \hat{L}}{\partial \boldsymbol{\lambda}_l}, \gamma$)
$\gamma_{t+1}$ = **UpdateLearningRate**($\gamma_t, t$)

We also selected the input word size of the RCCM accordingly. In the forward path, we approximate function $g$ in (2) with $G$, which uniformly quantizes a real number $x \in [0, m]$ to a $k$-bit number

$$G(x) = \frac{1}{2^f} \left\lfloor 2^f x + \frac{1}{2} \right\rfloor \quad (7)$$

where $\lfloor \cdot \rfloor$ returns the greatest integer less than or equal to the argument and $m$ is the upper bound. $m$ itself is bounded by its arbitrary unsigned 2's complement fixed-point representation where $f$ is the number of fractional bits and hence $m = 2^{k-f} - 2^{-f}$. A summary of the training process is given in Algorithm 1, which is similar to [3] and [6], with the addition of distribution matching and incorporating the quantization scheme of (5).

## VI. Experimental Setup

In this section, we present the system used to evaluate the benefits of our AddNet optimizations. We implemented the circuits in Figs. 2–4 in the hardware description language (HDL), very high speed integrated circuit HDL (VHDL), as they were more naturally described in an HDL than using other high-level synthesis tools. We then chose to integrate it into the open-source FPGA CNN Library by Alpha Data in VHDL [41], which provides basic neural network layers for generating custom 8-bit fixed-point CNN implementations. We instantiate the multiplier to replace the traditional VHDL fixed-point multiplication used in the original Alpha Data source code. This is used as our hardware library.

The bitstream generation workflow is illustrated in Fig. 6. After defining the CNN architecture and pretraining the network, the RCCM coefficients are calculated using distribution matching. The user provides this information to our AddNet tensorflow software library which then trains the network for a specified adder size. Once trained, the weights are written to a file. These weights, along with the parallelism factors and architectural preferences, are provided to the hardware library. The bitstream is then generated using Vivado 2018.1 with both the peripheral component interconnect express (PCIe) interface and network accelerator core, which are downloaded onto the FPGA. We tested both our tensorflow inference and hardware accelerator output to ensure correctness of the design.

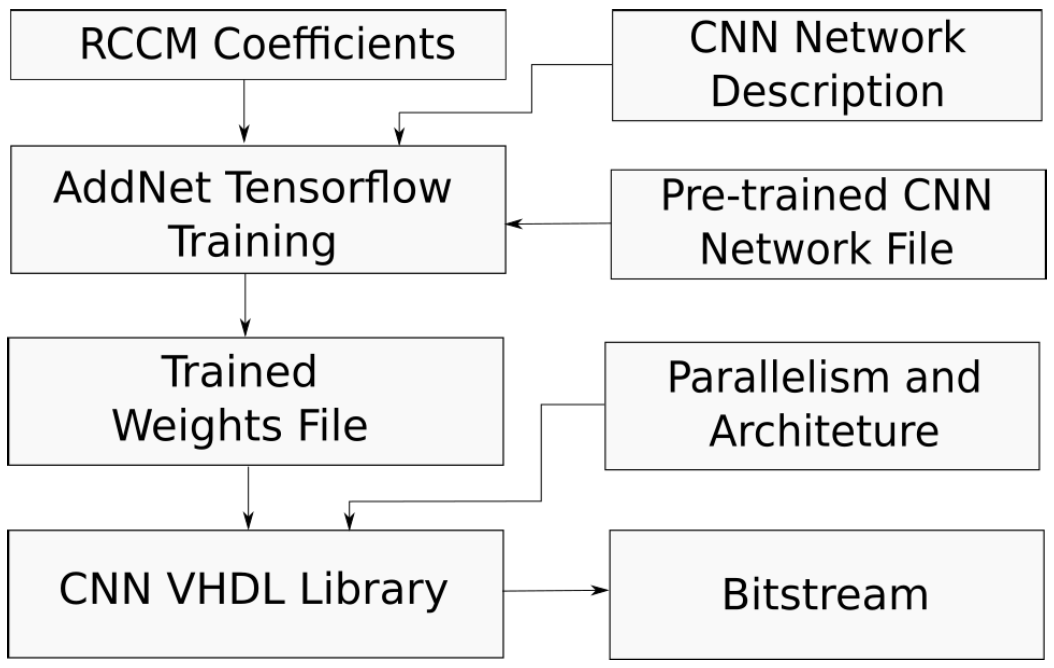

Fig. 6. Bitstream generation design flow.

### A. System Overview

The network accelerator core is integrated with a PCIe interface as illustrated in Fig. 7(a), which uses a streaming approach to direct memory access (DMA) data at an efficient rate across the PCIe bus and back. The design is targeted to the Alpha Data ADM-PCIe-8K5 board with a Xilinx KU115 FPGA, which consists of 2160 block random access memories (BRAMs) (36K), 5520 DSPs, and 663 000 LUTs. A board-specific PCIe Alpha Data IP core is used to interface between PCIe and our network accelerator core. This IP core can be configured to provide and consume AXI4 DMA streams of width 256 bits at a clock rate of 250 MHz in response to API function calls from the host. This stream width is reduced to match the buffer sizes for the inputs (24 bits) and weights (4 . . . 8 bits) which control data ingress. The weight data are sent in contiguous bursts to each layer in the network accelerator core to match the expected input behavior. DMA channel 0 is used to provide the input data for layer 0 and weights are initialized on DMA channel 1. The layer output is sent back over PCIe using a separate DMA channel. Additionally, a memory mapped direct slave port is used to access a bank of registers which can be read by the host to measure performance.

### B. Network Layer Accelerator Core

The network layer accelerator core performs the MAC operations in parallel to compute the convolution as in (1). The core receives input from the feature and weight buffers and writes to the output buffer. The feature and weight buffers stream data into a serial-to-parallel converter to fan-out the data to $n$ parallel processing elements (PEs). This is shown in Fig. 7(b). Once all data reach the PE, up to $p$ multiplications between features and weights are performed in parallel, and the results are accumulated. Here, we replace the standard 8-bit multiplier with our AddNet constant coefficient multiplier described in Section IV to reduce the cost per MAC over fixed-point implementations. The data are then accumulated before being fed into a ReLU activation function. The output then fans-in via a parallel-to-serial converter before being streamed out of the current layer and into the subsequent layer. After fanning in, the feature stream data are multiplied by an 8-bit scaling constant $\lambda_l$, which is precomputed. The number of multipliers is significantly larger than the number of layers in neural network designs. Hence, although we add one additional scale operation per layer, it only constitutes a tiny proportion of the overall area in comparison with high precision architectures.

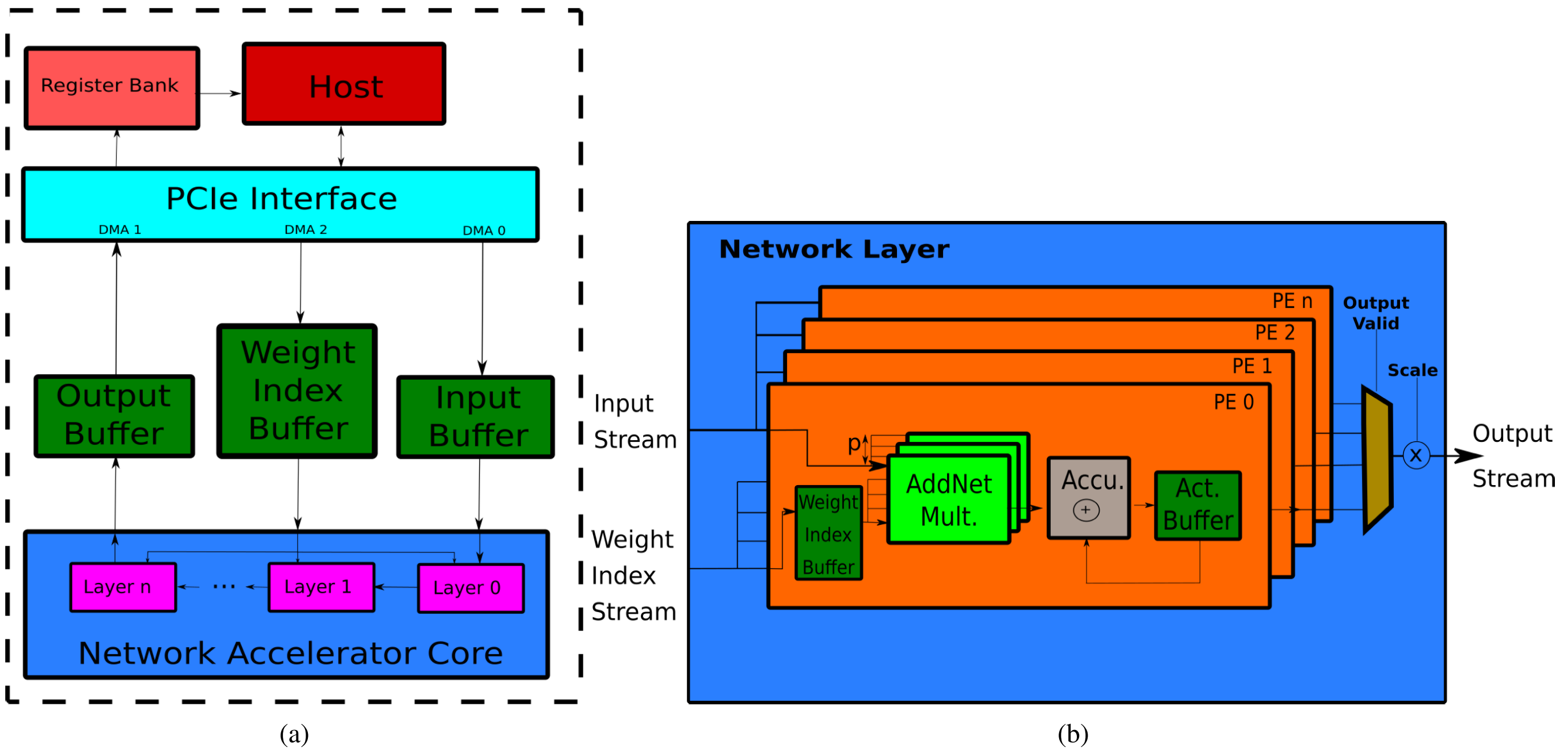

Fig. 7. Hardware accelerator system design. (a) AlphaData library system design. (b) Network layer accelerator core including AddNet multipliers.

### *C. Architectures*

To quantify the benefits of the AddNet optimizations, we use two different architectures with 2-Add, 3-Add, and 4-Add RCCMs, as well as traditional 8-bit fixed point. The architectures we study are a single-layer CNN accelerator and a full AlexNet-variant network [40], which reduces the filter size in the first layer to 7×7 and changes the stride of the first and second layers to 2. The single-layer implementation represents a loopback architecture. To implement a full network using this architecture, data are sent between the host and FPGA after each layer is computed sequentially. To minimize the amount of loopback iterations, we instantiate 2048 PEs as this equates to the number of neurons in the largest layer for all our networks. As such, we can compute all layer output feature maps for any of our networks during each loopback iteration. The AlexNet implementation represents a full dataflow where all convolutional layers are processed on the FPGA. For all AlexNet implementations using RCCMs, the first layer uses the 4-Add RCCM. This was because a higher number of coefficients was required in the first layer to achieve higher accuracy.

Both architectures were developed by AlphaData; we have not added any optimizations aside from our arithmetic operators. This enables us to focus on the benefits of our optimizations; we believe AddNet could improve on any 8-bit fixed-point deep-learning circuit.

### *D. Memory Use*

One important advantage of the RCCM designs is the reduction in the number of coefficients required for storage. Instead of storing the coefficients $c_s$, only the index $s$ has to be stored. This is similar to the weight-sharing approach. However, no decoder circuit is necessary to realize the codebook, as this is implicitly done by the proposed RCCM. While the coefficients in Table II would require 8, 10, and 12 bits to represent in 2's complement, storing the index requires only 4, 6, and 8 bits for the 2-Add, 3-Add, and 4-Add, respectively. So, significant savings in storage and memory bandwidth are possible for the 2-Add and 3-Add cases. For the Kintex Ultrascale devices, BRAMs can be a 36 K unit or 18 K units. As the number of 36 K BRAMs are reported, the weight buffers at each PE are calculated as 0.5 to represent 18 K BRAMs or 1 to represent 36 K BRAMs.

## VII. Results

We now display various hardware utilization and accuracy results to demonstrate applicability in neural network computation. The hardware results were obtained after place and route (PAR) using the Vivado 2018.1 design tool.

### *A. Reconfigurable Multiplier Resources*

First, we made a comparison of the resource usage of our proposed RCCM compared with a generic multiplier. As a generic multiplier, we selected the native Xilinx multiplier as it showed the best results for low word sizes [18]. Fig. 8 shows the LUT resources for varying input (activation) word sizes $w_{\text{in}}$ from 3 to 16 bits. While the generic multiplier grows at about 7.4 LUTs/bit, the proposed 2-Add, 3-Add, and 4-Add RCCMs only grow at 2, 3, and 4 LUTs/bit, respectively. It can be seen that the 2-Add and 3-Add RCCMs always outperform the generic multiplier for $w_{\text{in}} > 4$ bit, while the 4-Add RCCM is only interesting for larger word sizes of $w_{\text{in}} > 9$ bits. For the considered 9-bit activation, 55.2% and 32.8% of the LUTs can be saved by using the 2-Add and 3-Add RCCMs. This improves further as we increase the activation precision, suggesting that this multiplier and quantization method can be very effective for CNN inference applications and potentially on-chip neural network training; both these benefit from higher activation precision. For the multipliers used in this experiment, the pipelined RCCMs can operate between 350 MHz (4 bits) and 250 MHz (16 bits), while the generic multiplier can be clocked at between 200 MHz (4 bits) and 150 MHz (16 bits). Here, it is expected that the generic multiplier can be faster for faster timing constraints at the cost of additional resources.

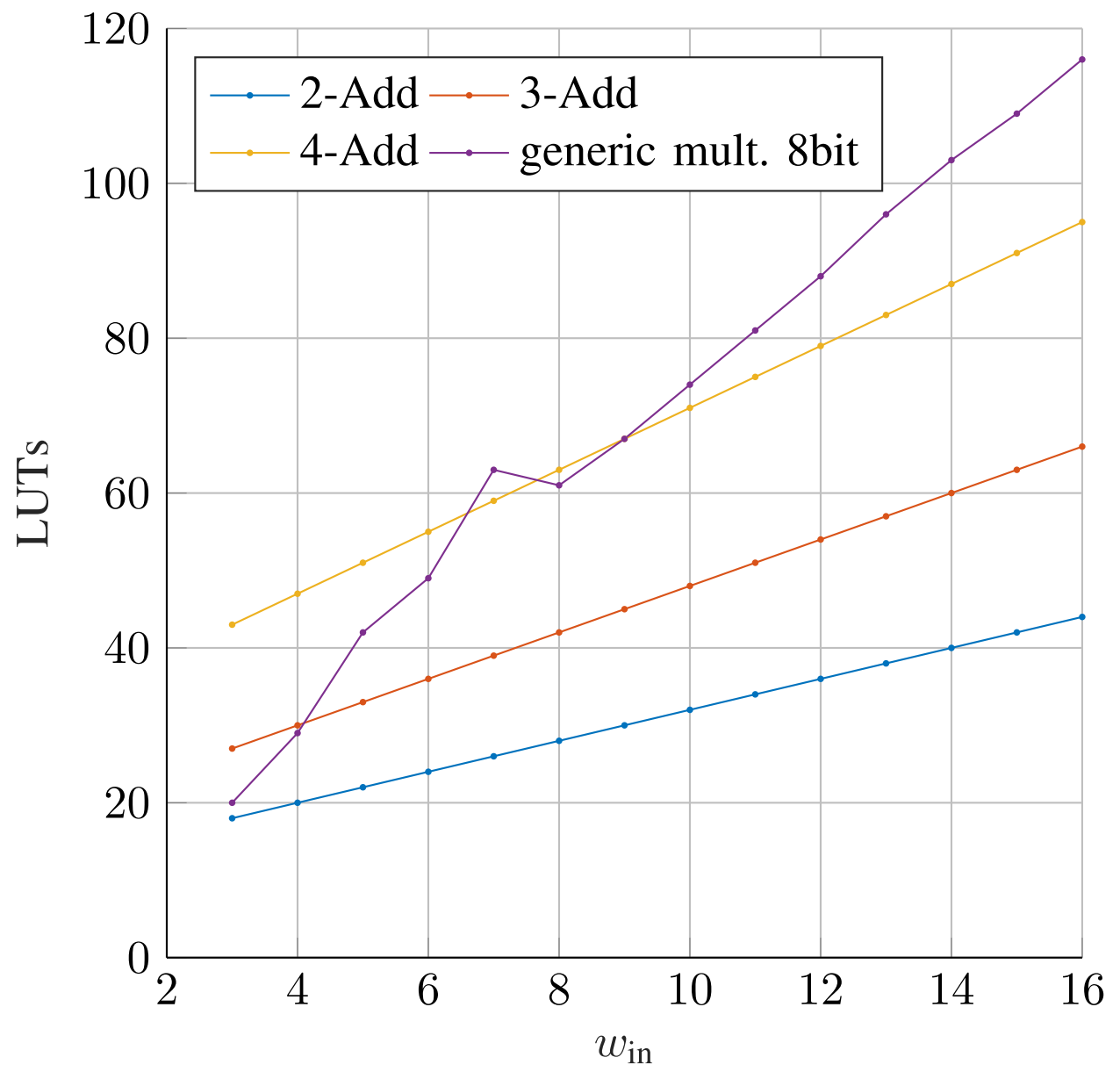

Fig. 8. LUT results from synthesis for the proposed RCCMs and a generic $8 \times w_{\text{in}}$ multiplier.

### *B. Architecture Resource Utilization*

In this section, we ran PAR experiments to compare our RCCMs against conventional 8-bit multipliers in a single CNN layer. Table V shows the resource utilization as well as the obtained speed.

The first row uses the fewest LUTs as multiplication is done in the DSPs. When DSPs are disabled, LUT usage dramatically increases. Our 2-Add design achieves the highest frequency at a significantly reduced LUT count compared with the 8-bit DSP-disabled implementation. However, we note that the LUT usage could be reduced if implemented with tree-structured

TABLE V

PAR Result Comparison of 1 Layer With Ten Neurons (With and Without DSP Mapping Enabled)

| Method | LUTs | FFs | DSPs | BRAM | Freq. [MHz] |
|---|---|---|---|---|---|
| 8-bit, with DSPs | 238 | 1015 | 10 | 5 | 446.63 |
| 8-bit, no DSPs | 1407 | 1515 | 0 | 5 | 355.11 |
| 2-Add | 818 | 1421 | 0 | 5 | 464.11 |
| 3-Add | 928 | 1365 | 0 | 5 | 415.15 |
| 4-Add | 1179 | 1385 | 0 | 5 | 342.68 |

TABLE VI

Summary of PAR Utilization on the Xilinx KU115 for All Arithmetic Types With PCIe Interface Included

| **Xilinx KU115** | **Architecture** | **2-Add** | **3-Add** | **4-Add** | **8-bit disab. [41].** | **8-bit enab. [41]** |
|---|---|---|---|---|---|---|
| BRAM (2160) | Conv Layer | 1154 | 1154 | 1154 | 1154 | 170 |
| | AlexNet | 1365 | 1557 | 1557 | 1557 | 1229 |
| DSP (5520) | Conv Layer | 48 | 48 | 48 | 48 | 96 |
| | AlexNet | 48 | 48 | 48 | 48 | 3760 |
| LUTs (663K) | Conv Layer | 187.0 | 205.6 | 255.8 | 383.0 | 36.2 |
| | AlexNet | 331.7 | 372.8 | 430.7 | 467.1 | 128.8 |
| Estim. Power | Conv Layer | 7.6W | 7.6W | 7.8W | 7.5W | 7.2W |
| | AlexNet | 39W | 44W | 48W | 52W | 29W |

TABLE VII

Per Layer BRAM Usage, $p$ Represents the Parallelism of the PE and $b$ Represents the Bits Required to Store Each Coefficient

| | $p$ | PE | b | | BRAMs | | Memory (MB) | |
|---|---|---|---|---|---|---|---|---|
| | | | 2-Add | 8-bit | 2-Add | 8-bit | 2-Add | 8-bit |
| Conv1 | 4 | 96 | 8 | 8 | 96 | 96 | 0.04 | 0.04 |
| Conv2 | 4 | 256 | 4 | 8 | 256 | 256 | 0.15 | 0.31 |
| **Conv3** | **1** | **384** | **4** | **8** | **192** | **384** | **0.44** | **0.88** |
| Conv4 | 2 | 384 | 4 | 8 | 384 | 384 | 0.33 | 0.66 |
| Conv5 | 2 | 256 | 4 | 8 | 256 | 256 | 0.44 | 0.88 |

optimizations as in [15]. The 3-Add and 4-Add designs have more flexibility compared with the 2-Add but require slightly more LUTs and operate with reduced frequency.

Table VI shows resource utilization for the two architectures, using the 2-, 3-, and 4-Add and 8-bit DSP-disabled designs. The 2-, 3-, and 4-Add cases all achieve significant LUT savings. The PCIe interface uses 48 DSPs and 100 BRAMs. Weight storage memory reduction is also apparent in the form of a decrease in BRAM utilization from 1557 in the 8-bit model to 1365 for 2-Add. This is because the reduction in bits per weight by using 2-Adders results in a 50% savings in BRAMs in the third convolutional layer, as highlighted in Table VII. This large savings is due to the discrete size of Xilinx BRAMs. Xilinx BRAMs can be configured to have a data width of 1, 2, 4, 9, and 18 bits. The wider the data width, the fewer number of words that can be stored per BRAM. The required data width for each PE is given by the bits stored per weight, $b$, multiplied by the parallelism of the PE, $p$. In the case of Conv3, where $p = 1$, by reducing $b$ to 4 bits, it is possible to store all the required weights for a PE using only a single 18 K BRAM, in contrast to a 36 K BRAM for the 8-bit case. In other layers where $p$ is higher, reduced memory use of two adders does not result in fewer BRAMs used due to their discrete sizes.

In Table VI, we also present the estimated overall board power. Evidently, there is an increase in board power when we disable DSPs. This is due to the increase of LUT resources that typically leads to more switching. However, comparing the 8 bits with DSPs disabled with AddNet implementations, we see reductions in power. Again, this is largely due to the reduction in LUTs.

The silicon area savings could also be used to scale up the parallelism to improve throughput, reduce latency, or fit the design on a smaller FPGA. For example, while the AlexNet and Conv Layer implementations already have one PE per output feature map, we can increase $p_l$ and compute more output feature map pixels in parallel to reduce the number of PE iterations required to compute a layer. This trivial optimization could be applied when accelerating most large neural networks. Typically such networks have lots of inherent parallelism and high computational requirements, with FPGA accelerators implementing some form of layer folding due to resource restrictions. This is especially the case for higher precision implementations when accuracy preservation is paramount. Thus, the AddNet multiplier is a very widely applicable tool for improving the parallelism of existing FPGA DNN architectures. Alternatively, these area reductions allow the current design to fit on a smaller device, such as the Xilinx VU3P, which would lead to expected reductions in power consumption as less hardware is being used.

### C. Frequency

The AlphaData CNN Library can operate conservatively at 250 MHz [41] and the critical path lies in the PCIe interface. Therefore, since our RCCMs can operate at a higher frequency, it does not translate to an increase in operating frequency of the overall system. However, as mentioned in Section IV-C, the RCCMs designed can also be trivially pipelined to improve their operating frequencies. As the multipliers were additionally implemented without the PCIe interface as both standalone components and within a CNN layer, we explored the frequencies of pipelined versions. The post-PAR frequencies with a clock constraint of 250 MHz (more aggressive time constraints will lead to higher frequencies than what is reported) for the standalone multipliers are shown in Table VIII. Significant frequency improvements are demonstrated from the pipelined versions which would lead to designs achieving higher frequencies when the multiplier lies in the critical path of the system. As the frequency is maintained at 250 MHz for all our implementations, the throughput remains constant.

### D. Effect of Layer Size

We now explore how the resource usage scales with parallelism for a single-layer convolutional core without the PCIe interface. This is an important metric for data flow implementations as we want to instantiate a higher number of PEs in layers with the most operations to achieve load balancing with less operationally intensive layers. Fig. 9 shows LUT usage from PAR where the number of parallel PEs is equal to typical neuron layer sizes used in our trained networks. Using such sizes allows us to simulate computing all output feature maps of a layer in parallel. As expected, all implementations scale linearly with the number of PEs. However, for the AddNet multipliers, as we increase the number of PEs, we see a smaller increase in LUTs in comparison with the 8-bit version. This is amplified further with the smaller multiplier implementations which demonstrate smaller gradients to the 4-Add version. For example, with 2048 PEs instantiated, we achieve a substantial 52% LUT reduction. Typically neural network implementations are constrained by the number of PEs we can instantiate per layer due to resource scaling.

TABLE VIII

PAR FREQUENCIES FOR PIPELINED VERSIONS OF THE RCCMS

| Type | 2-Add | 3-Add | 4-Add |
|---|---|---|---|
| Original | 447.43 MHz | 483.09 MHz | 342.82 MHz |
| Pipelined | 770.42 MHz | 578.03 MHz | 623.83 MHz |

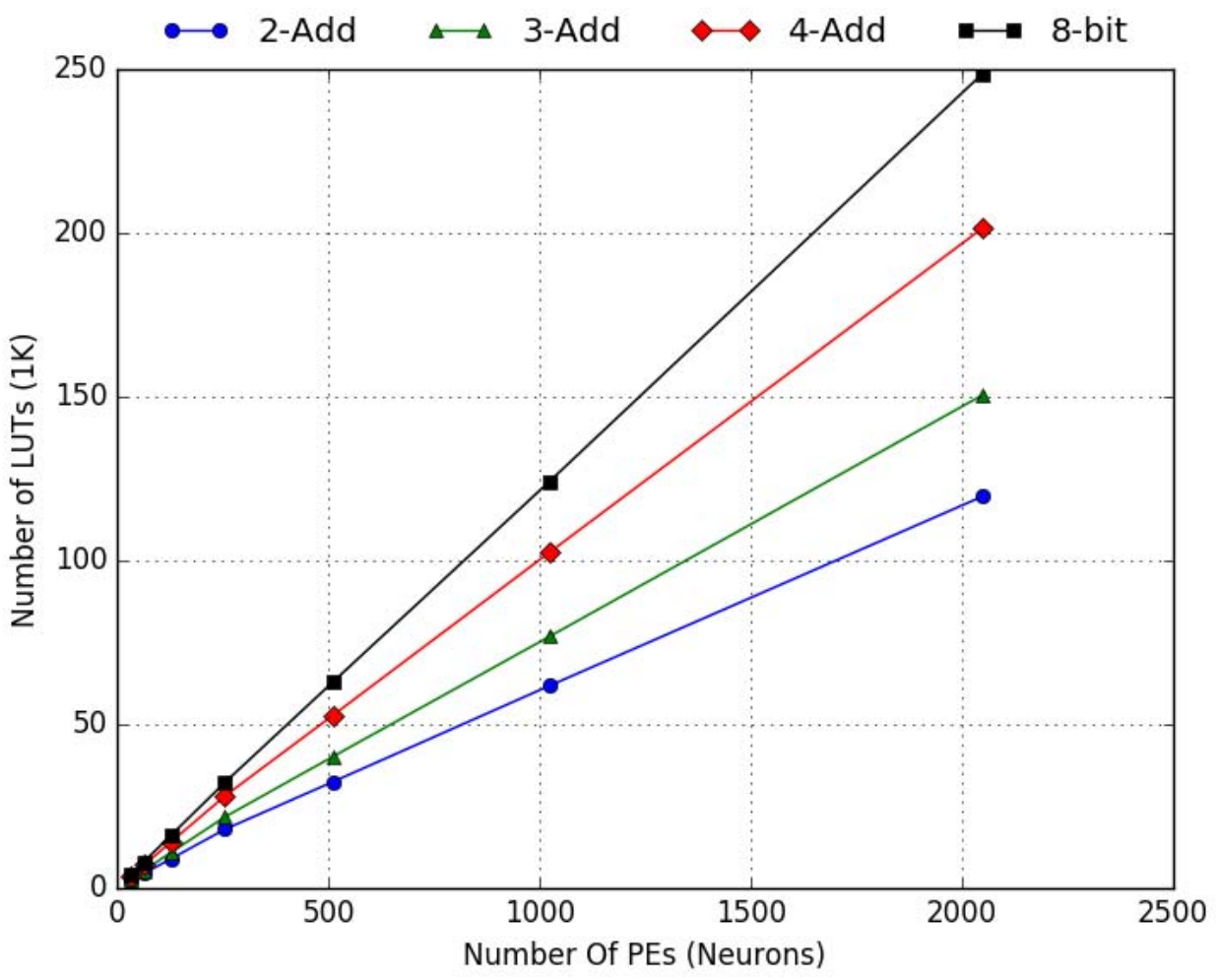

Fig. 9. Relationship between LUTs and amount of parallelism for different arithmetic operations.

### E. Accuracy

To demonstrate the robustness of our quantization strategy, we implement the training on several benchmark networks for image classification. The proposed method is evaluated on the ILSVRC-2012 ImageNet data set which contains natural high-resolution visual classification data set consisting of 1000 classes, 1.28 million training images, and 50 K validation images. The images are preprocessed as per the reference models by resizing the inputs to 256×256 before being randomly cropped to 224×224. We report our single-crop performance evaluation results using Top-1 and Top-5 accuracy, where the cross-entropy loss of the predicted classification against the actual classification is minimized during training. The AlexNet network consists of five convolutional and three fully connected layers. ResNet networks consist of blocks of two or three convolutional layers and a residual connection [11]. Two models are explored with varying depths of these blocks.

TABLE IX
ACCURACY RESULTS (%) FOR ADDNET, FLOATING-POINT (32-bit) TRAINING, AND FIXED-POINT TRAINING OVER VARIOUS IMAGENET MODELS

| **Model** | | **2-Add** | **3-Add** | **4-Add** | **float.** | **8-bit** | **6-bit** | **4-bit** | **Ternary** | **Binary** |
|---|---|---|---|---|---|---|---|---|---|---|
| AlexNet | Top-1 | 55.8 | 55.8 | 55.9 | 55.1 | 55.5 | 54.7 | 53.9 | 53.2 | 52.0 |
| | Top-5 | 79.8 | 79.8 | 80.0 | 79.2 | 78.6 | 78.5 | 78.3 | 78.1 | 76.9 |
| ResNet-18 | Top-1 | 65.1 | 66.0 | 66.4 | 68.6 | 66.0 | 63.5 | 62.0 | 61.6 | 57.5 |
| | Top-5 | 86.4 | 87.6 | 87.8 | 88.2 | 87.5 | 85.9 | 85.4 | 84.2 | 81.2 |
| ResNet-50 | Top-1 | 72.1 | 72.7 | 73.3 | 76.0 | 72.5 | 69.6 | 68.4 | 67.0 | 65.0 |
| | Top-5 | 91.2 | 91.5 | 92.0 | 92.9 | 91.6 | 89.5 | 89.1 | 88.7 | 86.5 |

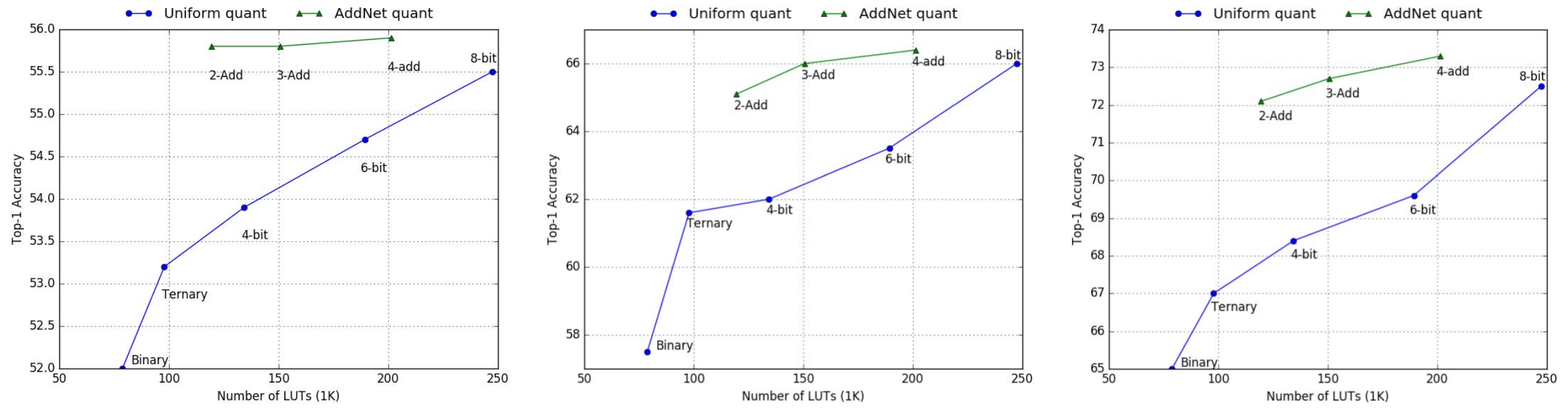

(a) (b) (c)

Fig. 10. Accuracy–area comparison of uniform and AddNet quantizations for AlexNet and ResNet. (a) AlexNet. (b) ResNet-18. (c) ResNet-50.

In Table IX, we display the accuracies of quantizing for different multiplier sizes and compare them with fixed-point retraining and floating-point network accuracies. All results were trained with the 4-Add RCCM in the first and last layers to preserve accuracy and were trained for a fixed number of epochs. For all these networks, we achieve at least 8-bit accuracy with resource savings through our multiplier. This demonstrates the effectiveness of AddNet. In particular, we can achieve equivalent to floating-point accuracy for AlexNet with only 2-Add multipliers, which translates to large resource savings. In some instances, the accuracy is improved and this is due to the regularization effect of the quantization which improves the generalization of the network.

### *F. Accuracy Versus Area*

Fundamentally, our goal is to achieve the highest possible accuracy while consuming the smallest amount of resources. Thus, it is important to evaluate the accuracy achieved against the amount of resources used. To do this, we have analyzed the area consumed for different precisions of traditional fixed-point training against AddNet training. Fig. 10 shows these evaluations for each network. The closer the data points are to the top-left corner of the graphs, the more optimized and more efficient the method. We see that both the 2-Add and 3-Add cases show improvements over the traditional quantization methods. This demonstrates the effectiveness of our training methodology. The 4-Add case achieves the same or greater accuracy than the 8-bit implementation but with significantly less resources. Additionally, for all the three networks, the 2-Add and 3-Add cases significantly improve accuracy and area over 6-bit implementations, and the 2-Add case significantly improves accuracy and area over the 4-bit implementations. This is a very important contribution of this work: instead of reducing precision, which is a standard approach to save silicon area, our method gets better area savings and much better accuracy for all networks.

TABLE X
ACCURACY RESULTS OF CHANGING ACTIVATION BIT WIDTH FOR 2-ADD RESNET-18

| | 2-bit | 4-bit | 8-bit | 12-bit | 16-bit |
|---|---|---|---|---|---|
| Top-1 | 54.3% | 57.7% | 58.2% | 58.2% | 58.2% |
| Top-5 | 79.8% | 81.4% | 82.0% | 82.0% | 82.0% |

After investigating the effect of weight precision, we also analyze the effect of activation precision on both accuracy and area. Table X shows the accuracy against different sizes for $w_{\text{in}}$ using the 2-Add multiplier coefficients for ResNet-18. We particularly analyze 2-Add ResNet-18 as it has the highest discrepancy from our 8-bit and full-precision models. As shown, increasing the activation to 16 bits does not close the accuracy gap. Observing Fig. 8, the area of the 2-Add with $w_{\text{in}} = 16$ is roughly equivalent to the 3-Add with $w_{\text{in}} = 8$. Thus, in this case, it is much more effective to use the 3-Add with $w_{\text{in}} = 8$ as the accuracy is improved.

## VIII. Conclusion

In this article, we explored reconfigurable constant coefficient multipliers for the inference problem in CNNs. A novel distribution matching scheme that restricts the allowable coefficient values in a computationally tractable manner and a training algorithm are proposed. Our results show that this approach achieves better accuracy than networks that constrain weight parameters to have binary or ternary values, while allowing the expensive multipliers usually used in fixed-point implementations to be replaced by shifters, adders, and small MUXes. Furthermore, the restricted number of possible coefficient values allows an encoding scheme to significantly reduce weight storage. Overall, our approach reduces mismatch between CNN computation and existing FPGA device architectures, making more efficient implementations possible.

While we have demonstrated the benefits of these techniques on CNNs, we expect that training any type of neural network to make use of RCCMs using our approach should be explored as an alternative to simply studying the use of reduced precision arithmetic. Our technique introduces a new dimension for the optimization of neural networks in which the arithmetic is directly customized and is orthogonal to matrix decomposition and sparsity-inducing approaches. Future work will explore combining these techniques.

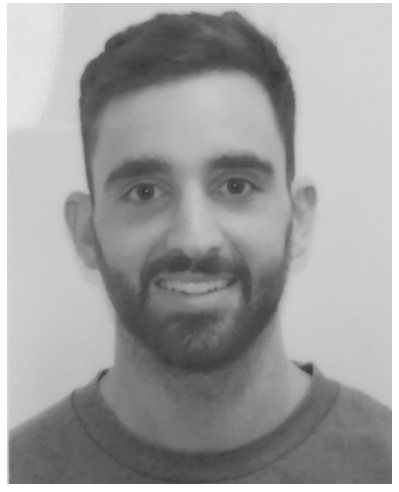

**Julian Faraone** received the B.Com. degree in quantitative finance and the B.E. degree in electrical and electronic engineering from The University of Western Australia, Crawley, WA, Australia, in 2015. He is currently working toward the Ph.D. degree at the Computer Engineering Laboratory, The University of Sydney, Sydney, NSW, Australia. During his Ph.D., he has interned at Xilinx Research Labs, San Jose, CA, USA, and Mythic-AI, Redwood City, CA, USA.

His current research interests include low-precision deep learning, field-programmable gate arrays, and optimization.

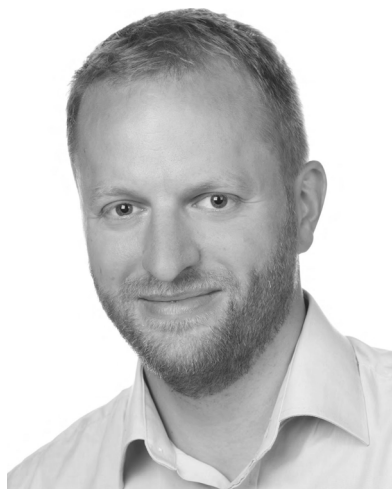

**Martin Kumm** (M'13) received the Dipl.-Ing. degrees in electrical engineering from the Fulda University of Applied Sciences, Fulda, Germany, and the Technical University of Darmstadt, Darmstadt, Germany, in 2003 and 2007, respectively, and the Ph.D. (Dr.-Ing.) degree from the University of Kassel, Kassel, Germany, in 2015.

From 2003 to 2009, he was at GSI Darmstadt, Darmstadt, where he was involved in digital RF control systems for particle accelerators. He is currently a Professor of Embedded Systems at the Fulda University of Applied Sciences. His current research interests include arithmetic circuits and their optimization and high-level synthesis, all in the context of reconfigurable systems.

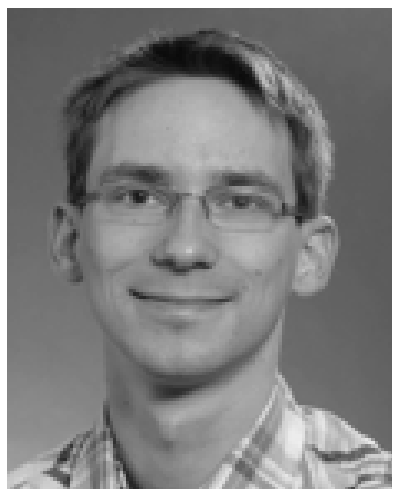

**Martin Hardieck** received the Dipl.-Ing. and M.Sc. degrees in electrical engineering from the University of Kassel, Kassel, Germany, in 2014 and 2016, respectively, where he is currently working toward the Ph.D. degree at the Digital Technology Group.

His current research interests include digital signal processing, model-based design, and artificial neural network implementations, all in the context of field-programmable gate arrays.

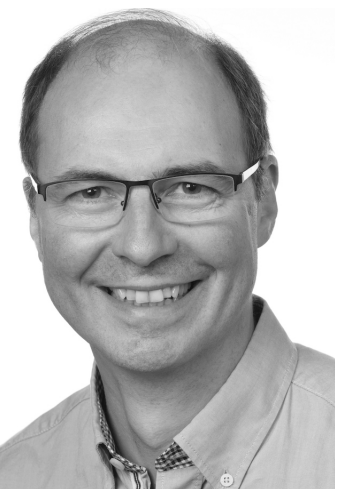

**Peter Zipf** (M'05) received the B.A. degree in computer science from the University of Kaiserslautern, Kaiserslautern, Germany, in 1994, and the Ph.D. (Dr.-Ing.) degree from the University of Siegen, Siegen, Germany, in 2002.

He was a Postdoctoral Researcher at the Department of Electrical Engineering and Information Technology, Darmstadt University of Technology, Darmstadt, Germany, until 2009. He is currently the Chair of Digital Technology at the University of Kassel, Kassel, Germany. His current research interests include reconfigurable computing, embedded systems, and system-on-chip designs, and design methodologies and CAD algorithms for circuit optimization and reconfigurable systems.

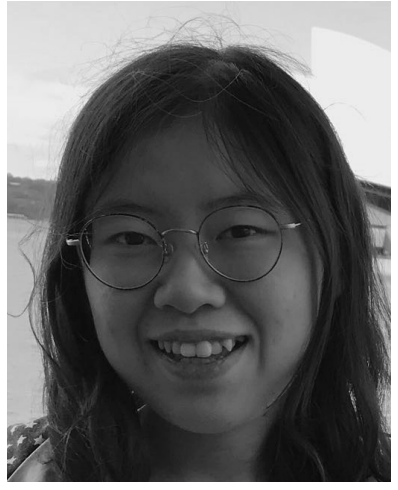

**Xueyuan Liu** received the B.E. degrees (Hons.) in electrical engineering from The University of Sydney, Sydney, NSW, Australia, and the Harbin Institute of Technology, Harbin, China, in 2019.

Her current research interests include deep learning and low-latency field-programmable gate array (FPGA) architecture.

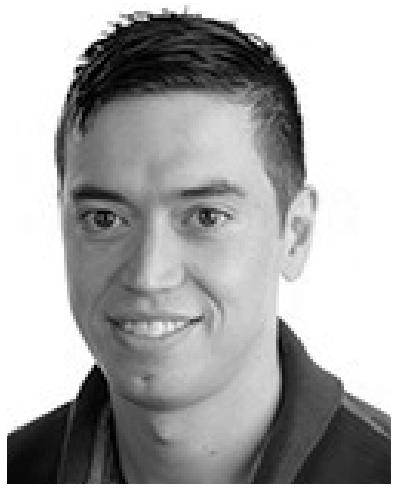

**David Boland** received the M.Eng. degree (Hons.) in information systems engineering and the Ph.D. degree from Imperial College London, London, U.K., in 2007 and 2012, respectively.

From 2013 to 2016, he was at Monash University, Melbourne, VIC, Australia, as a Lecturer before moving to The University of Sydney, Sydney, NSW, Australia. His current research interests include numerical analysis, optimization, design automation, and machine learning.

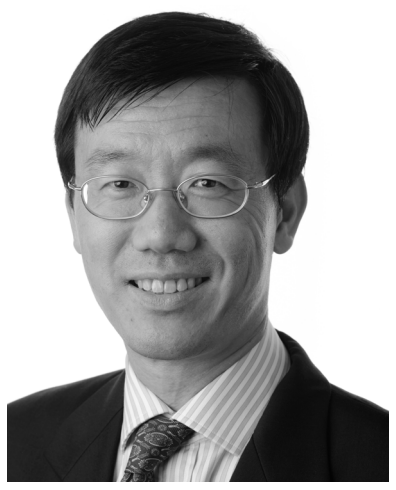

**Philip H. W. Leong** (SM'02) received the B.Sc., B.E., and Ph.D. degrees from The University of Sydney, Sydney, NSW, Australia, in 1987, 1989, and 1993, respectively.

In 1993, he was a Consultant at ST Microelectronics, Milan, Italy. From 1997 to 2009, he was with The Chinese University of Hong Kong, Hong Kong. He is currently a Professor of Computer Systems at the School of Electrical and Information Engineering, The University of Sydney, a Visiting Professor at Imperial College London, London, U.K., a Visiting Professor at the Harbin Institute of Technology, Harbin, China, a Chief Technology Advisor at ClusterTech, Hong Kong, and a Visiting Scholar at Xilinx Research Labs, San Jose, CA, USA.